\documentclass[%
 aip,
 amsmath,amssymb,
 reprint,%
]{revtex4-1}

\usepackage{graphicx}
\usepackage{dcolumn}
\usepackage{bm}
\usepackage{afterpage}

\usepackage[utf8]{inputenc}
\usepackage[T1]{fontenc}
\usepackage{mathptmx}
\usepackage{etoolbox}
\usepackage{caption}
\usepackage{subcaption}

\makeatletter
\def\@email#1#2{%
 \endgroup
 \patchcmd{\titleblock@produce}
  {\frontmatter@RRAPformat}
  {\frontmatter@RRAPformat{\produce@RRAP{*#1\href{mailto:#2}{#2}}}\frontmatter@RRAPformat}
  {}{}
}%
\makeatother
\begin{document}

\preprint{AIP/123-QED}

\title[]{Wavenumber Calibration for an Imaging Refractometer}
\author{A. Rososhek}
    \email{ar877@cornell.edu}
\author{B. R. Kusse}
\author{W. M. Potter}
\author{N. J. Wilson}
\author{E. S. Lavine}
\author{D. A. Hammer}
\affiliation{Laboratory of Plasma Studies, Cornell University Ithaca, New York 14853, USA
}%
 \homepage{https://www.lps.cornell.edu}

\date{\today}

\begin{abstract}
An imaging refractometer can be used to describe the properties of a high-energy density plasma by analyzing the transverse intensity distribution of a laser beam that has passed through the plasma. The output of the refractometer can be directly calibrated in terms of beam deflection angles using ray transfer matrix analysis. This paper describes a novel way to calibrate the refractometer output in terms of the spatial wavenumbers of the transverse intensity distribution of the laser beam. This is accomplished by replacing the plasma with a gridded structure that modulates the transverse intensity of the beam, producing an intensity distribution with a known Fourier Transform. This calibration technique will generate a one-to-one mapping of deflection angle to wavenumber and will enable measurement of the size of Fourier space available to the system. The spectrum of wavenumbers generated when the laser beam passes through a high-energy density plasma may contain information about the types of density fluctuation that are present in the plasma.
\end{abstract}

\maketitle

\section{\label{sec:intro}Introduction\protect
}
\par The power spectrum of density fluctuations has numerous applications in various fields, including high-energy density (HED) plasmas, cosmology, and astrophysics. The power spectrum contains information about the density and energy distribution on different scales that can be used to analyze, for example, the evolution of the universe \cite{schneider1995power} and gamma-ray bursts.\cite{mesinger2005constraints} In astrophysics, the power spectrum is utilized in a variety of studies, including those focused on large and small scale turbulence in the interstellar medium \cite{armstrong1981density}, electron density fluctuations in the solar wind,\cite{woo1979spacecraft} and turbulence in the magnetosheath of Saturn.\cite{hadid2015nature} Typically, the power spectrum of density fluctuations is inferred from spectroscopy data, with one of the most prominent data sources being the 21 cm line of hydrogen, among others.\cite{loeb2004measuring,barkana2005method,loeb2008possibility} In HED plasmas, such as those produced at the National Ignition Facility, X-ray Thomson scattering can be utilized \cite{glenzer2009xray} to obtain information about the fluctuating density. In other experimental settings, the wavenumber spectra can be obtained via different imaging techniques like schlieren or shadowgraphy, where laser light intensity measured at the detector screen reflects the first (schlieren) and second (shadowgraphy) derivative of the refractive index. These methods require careful post-processing of the image, as shown in Ref. [\onlinecite{white2019supersonic}], where the use of digital Fourier Transform (FT) schemes is likely to introduce numerical noise. Such techniques can still be challenging to interpret when density gradients cause ray crossing and caustics. Another way to interpret the schlieren imaging data is to adopt the so-called quantitative schlieren technique where the apparatus employs a suitable light source and the deflection angle calibration is done via a calibration lens.\cite{calibr_schlieren} An example of a theoretical approach that establishes a relation between the FT of the measured schlieren intensity and the power density spectrum is given in Ref. [\onlinecite{white2019supersonic}]. The authors believe the imaging refractometer is a variation of schlieren imaging with built-in optical FT capability. Therefore, this theoretical relation applies to it as well.
\par A recently reformulated laser wavefront probing diagnostic from Ref. [\onlinecite{ascolibartoli1965plasma}], developed in [\onlinecite{hare2021imaging}], enables the measurement of deflection angles along one axis and provides spatial information along the other. This instrument, an imaging refractometer, is sensitive to variations in the refractive index in plasma and has a high dynamic range. The technique is particularly useful in symmetric configurations, for example, planar and cylindrical geometries, and quasi-symmetric (symmetry is perturbed) environments, where averaging across the symmetry axis can increase the signal-to-noise ratio. 
In this diagnostic, the deflection angle is generated through the refraction of the incoming collimated laser beam, preferably with a single spatial mode. This deflection angle represents the angle between the perturbed and unperturbed rays and is proportional to the integral of the refractive index gradient over the optical path shown in Eq. \ref{def_angle}.\cite{hutchinson} To interpret the spectrum of deflection angles as being related to the power spectrum of density fluctuations, it is crucial to establish a connection between deflection angle and wavenumber, \emph{k}. The deflection angle calibration can be achieved by solving the ray transfer matrix equations,\cite{hare2021imaging} which provide a linear relation between the pixel locations on the screen and deflection angles. A separate establishment of a linear pixel-to-wavenumber relation implies a linear connection between deflection angles and wavenumbers (at the Fourier Plane) as well. In this paper, we propose a novel method to form such a linear correspondence between pixels and wavenumbers by calibrating the optical system with an object plane target whose FT is analytically known. For this purpose, a family of Ronchi rulers,\cite{Ronchi:64} which are typically defined by a transparent-opaque line pair (lp) density in lp/mm, can be used. The target used in experiments for the current paper has 2 lp/mm, meaning that the size of a transparent line is 250 $\mu m$. This size along with the 532 nm laser used in this work implies that laser wavelength-size-related diffraction has almost no effect on the pattern formed by the imaging refractometer.
\begin{eqnarray}
    \theta = \frac{1}{2} \int \frac{\nabla n_e}{n_{cr}} dl \label{def_angle}
\end{eqnarray}
\par To avoid possible ambiguities, it is important to address the relationship between wavenumbers observed at the Fourier plane and those within the plasma. The application of the calibration presented here may be challenging in the presence of sufficiently thick plasmas with small-scale density perturbations and turbulence where the probing beam can be imagined as a random walk.\cite{Merlini_2023} This leads to a Markov property, which implies a memoryless process that complicates the determination of the specific probability density that generated the distribution. However, a random walk process involves phase randomization and leads to formation of a speckle pattern, which carries information that may be available via intensity and speckle interferometry.\cite{brown1956,labeyrie1970attainment, Goodman1975} The theoretical and experimental thresholds where laser propagation through turbulent plasmas should or should not be treated as a random walk and the applicability of the intensity/speckle interferometry are outside the scope of this paper and will be discussed elsewhere.
\begin{figure}[h]
    \includegraphics[width=\linewidth]{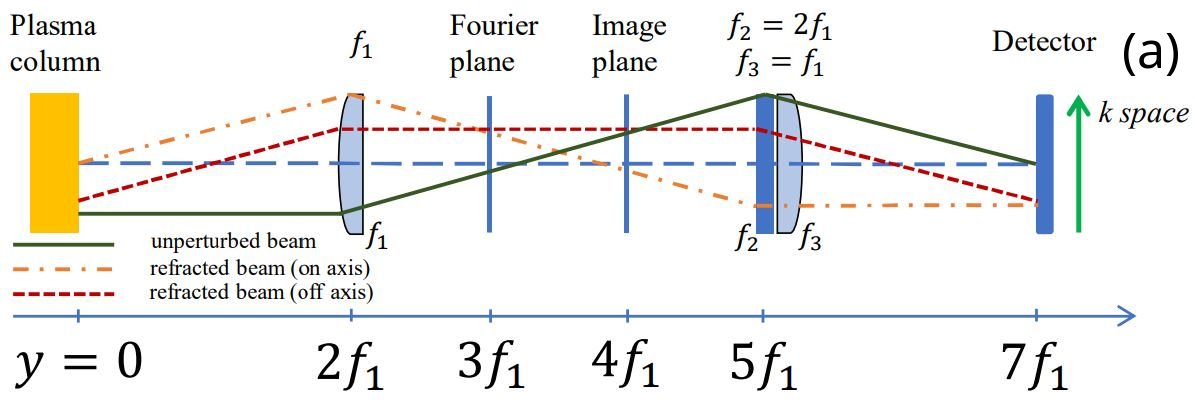}
    \includegraphics[width=\linewidth]{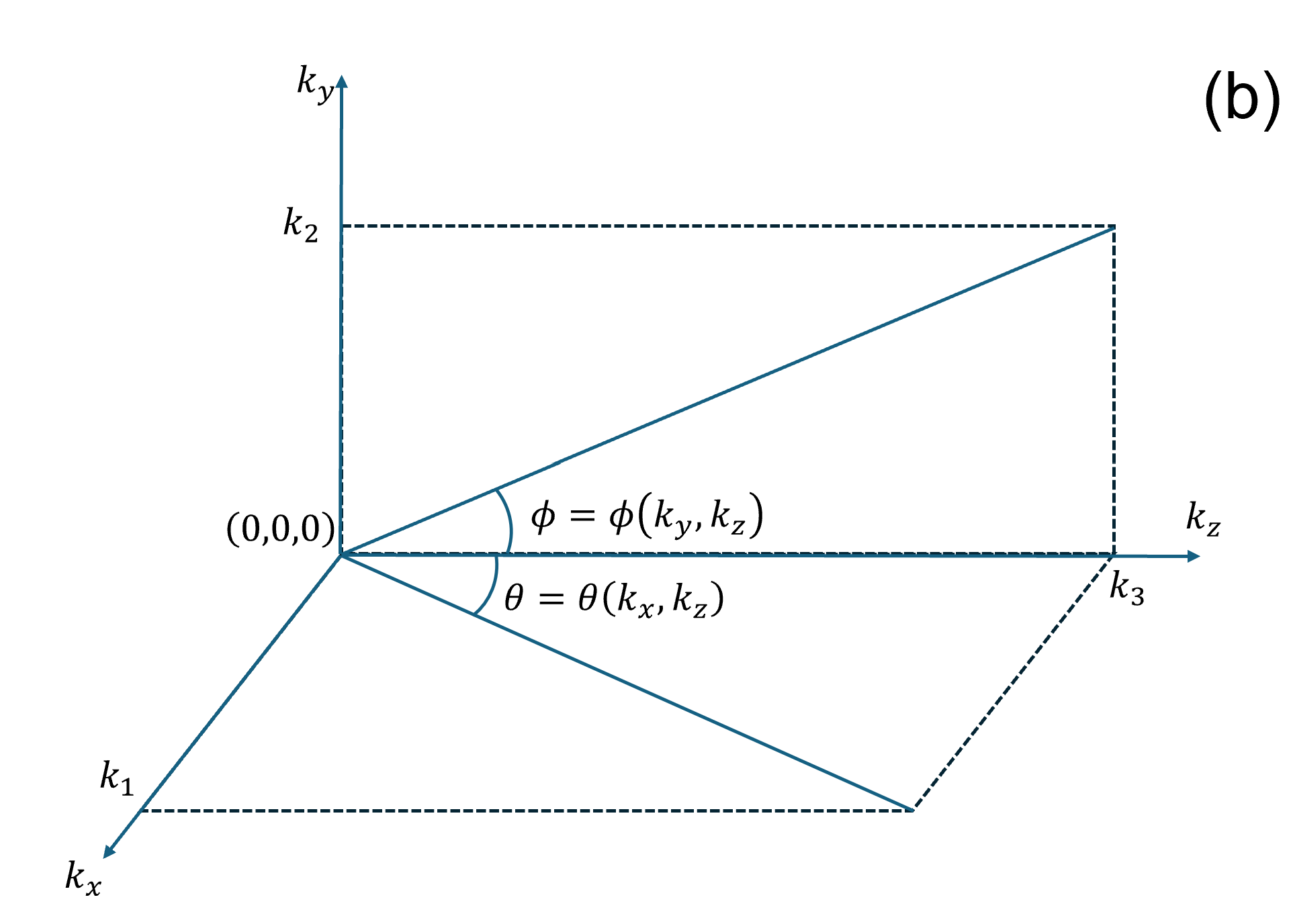}
    \caption{\label{opt_scheme} (a) A schematic view of the optical setup for an imaging refractometer showing the optical Fourier transforming (the first) lens and the compound lens (cylindrical and spherical back-to-back); (b) a geometrical representation of the projections of the wave vector $\mathbf{k}=k_1\hat{k}_x+k_2\hat{k}_y+k_3\hat{k}_z$ onto $k_x-k_z$ and $k_y-k_z$ planes.}
\end{figure}
\section{\label{sec:background}Mathematical background\protect
}
In what follows, the connection will be established between the Fourier-transformed density distribution at the object plane with the measurement taken from the Fourier plane of the optical setup (see Fig. \ref{opt_scheme}a). 

\par Before proceeding with the description of the physical picture, let us define the main functions and variables involved. The real-space is the three-dimensional (3D) $\mathit{(x,y,z)}$ space, where $\mathit{z}$ will be the laser propagation direction, and the $\mathit{x-y}$ plane will contain the transverse spatial distribution of a laser field. The spatial frequency domain will then be formed by $\mathit{(k_x, k_y, k_z)}$ variables. The laser is incoming from $\mathit{z=-\infty}$, its propagation direction is along the $\mathit{z}$-axis, and we assume it is distributed on the $\mathit{x-y}$ plane only. In the Fourier space, one can speak of the propagation direction in terms of angles, invoking the so-called angular spectrum framework.\cite{goodman} There, the laser field is described as a superposition of plane waves with each traveling in a direction set by $(\theta(k_y,k_z),\phi(k_x,k_z)$. In Fig. \ref{opt_scheme}b, we show the two projections onto $k_x-k_z$ and $k_y-k_z$ planes of a wave vector $\mathbf{k}$ that represents one specific plane wave. Here, one may notice that if the change in $k_z$ is sufficiently small, then $\theta$ and $\phi$ angles become directly related to $k_y$ and $k_x$ variables respectively. Moreover, note that these angles are projection angles from the general direction of propagation $\mathbf{k}$ onto the $k_y-k_z$ and $k_x-k_z$ planes, while the deflection angle defined via Eq. \ref{def_angle} is an angle between $\mathbf{k}$ and its initial direction before entering the interaction region, which is ${\hat{k}}_z$.
\par The intensity of the laser field is represented in real-space by $A_0(x,y) = C_0 \times h(x,y)$, where $\mathit{h(x,y)}$ is the transverse spatial intensity distribution in the $\mathit{x-y}$ plane and $C_0$ is the amplitude of the laser field. In the Fourier space, the FT of the incoming laser field intensity $A_0$ will take the following form: $\widetilde{A}_0(k_x,k_y)=\mathcal{F}[C_0 \times h(x,y)](k_x,k_y)$. It is worth mentioning here that if the laser field has a Gaussian form (single TEM\textsubscript{00} mode laser profile), then its FT will also be a Gaussian with most of the laser energy being concentrated around $(k_x,k_y)=(0,0)$ of the Fourier plane. Additionally, the plasma region is centered at $\mathit{z=z_0}$, which we will call the object plane, and is assumed to have width $\delta z$ along $\mathit{z}$-axis direction, which ideally would be negligibly small ($\delta z\rightarrow0$). Furthermore, let us assume that the FT of the plasma density distribution exists in the immediate neighborhood of $z=z_0$ and will be denoted as $\widetilde{n}(k_x,k_y,k_z)$. Here we invoke the concept of an optical transfer function\cite{goodman} (OTF) that will contain all the information about the plasma density distribution and its effect on the intensity (modulation transfer function, MTF) and phase (phase transfer function, PTF) of the laser field. Namely, the MTF will include the information about the plasma distribution at $z=z_0$ and it will act to modulate the incoming laser field intensity according to this distribution. Let us denote the MTF by $\mathcal{M}(k_x,k_y,k_z)$ and the PTF by $\Phi(k_x,k_y,k_z)$ such that the OTF is defined via Eq. (\ref{otf}).
\begin{eqnarray}
    \mathcal{H}(k_x,k_y,k_z) = \mathcal{M}(k_x,k_y) \times \exp(\mathrm{i} \Phi(k_x,k_y))  \label{otf}
\end{eqnarray}
\par We start with a single-mode (TEM\textsubscript{00}) laser beam incident from the $-z$ direction onto a plasma region at $z=z_0$, i.e., the intensity distribution is given by $A_0$. In the Fourier space, the resultant laser field intensity immediately after the plasma region ($z>z_0$ in real space) will be given via Eq. (\ref{res_field}).
\begin{eqnarray}
    \left| \widetilde{A}(k_x,k_y) \right|^2 = \left| \widetilde{A}_0(k_x,k_y) \times \mathcal{M}(k_x,k_y) \right|^2 \label{res_field}
\end{eqnarray}
Therefore, the resultant laser field intensity that was concentrated around $(k_x=0,k_y=0)$ will now be redistributed to the wavenumbers present in $\widetilde{n}(k_x,k_y,k_z)$.  A lens placed, for example, two focal length away (see Fig. \ref{opt_scheme}a) with the plasma at its object plane, will form Fourier and Image planes. At the Fourier plane, the FT of the resultant laser field will appear; the modulated laser field intensity will be seen at the Image plane. It is clear, that if the plasma density is negligibly small between the object plane and the lens then to obtain the intensity distribution at the Fourier plane formed by the lens, one would need to multiply $\widetilde{A}$ by an appropriate propagator in Fourier space and another function, the OTF of the lens that specifies how a light field would propagate through the lens. Also, it is helpful to invoke Babinet's principle\cite{goodman}, which states that apertures produce identical intensity patterns (apart from the origin) in the far field, or at the Fourier plane in our case, as the same size and shape obstacles do. One might see that as an equivalence between the regions where plasma density is beyond critical with transparent regions of the same shape so that both will be present in the spatial frequency spectrum. The wavenumber corresponding to a specific shape present within the plasma density distribution will have a non-zero intensity at the Fourier plane. Therefore, we conclude that for the simplest case of infinitely thin plasma density at the object plane, its 2D FT amplitude, a power spectrum, is measurable at the Fourier plane of the lens. Furthermore, hybrid imaging by an imaging refractometer will result in a measurement of either $\theta$ or $\phi$ defined in Fig. 1(b), which is a component of the FT plasma density distribution along with its real-space x-component at the object plane. Such a conclusion also leads to the possibility of measuring the full 2D FT of the plasma density distribution by adding an additional imaging refractometer leg that would have its cylindrical lens rotated 90 degrees with respect to the first leg. This arrangement would also require careful pixel-to-pixel alignment of the two legs. Another conclusion is that the density distribution outside of the object plane but within the optical setup depth of field will add another factor into the product shown in Eq. \ref{res_field}. 

    \begin{figure*}[t]
     \centering
     \begin{tabular}{cc}
        \hspace*{\fill}%
      \includegraphics[width=.3\paperwidth]{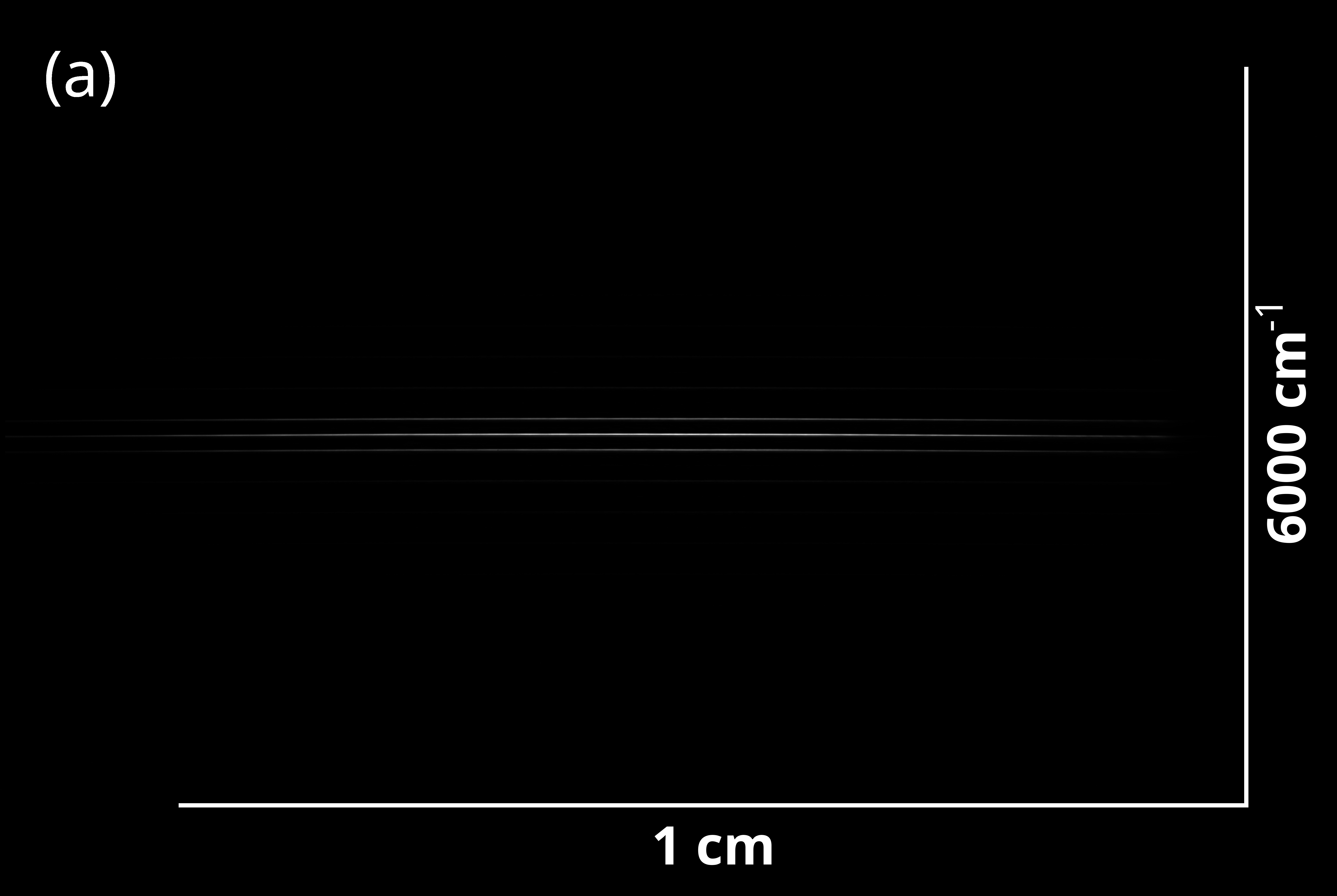} \hspace{0.02\paperwidth}
      \includegraphics[width=.35\paperwidth]{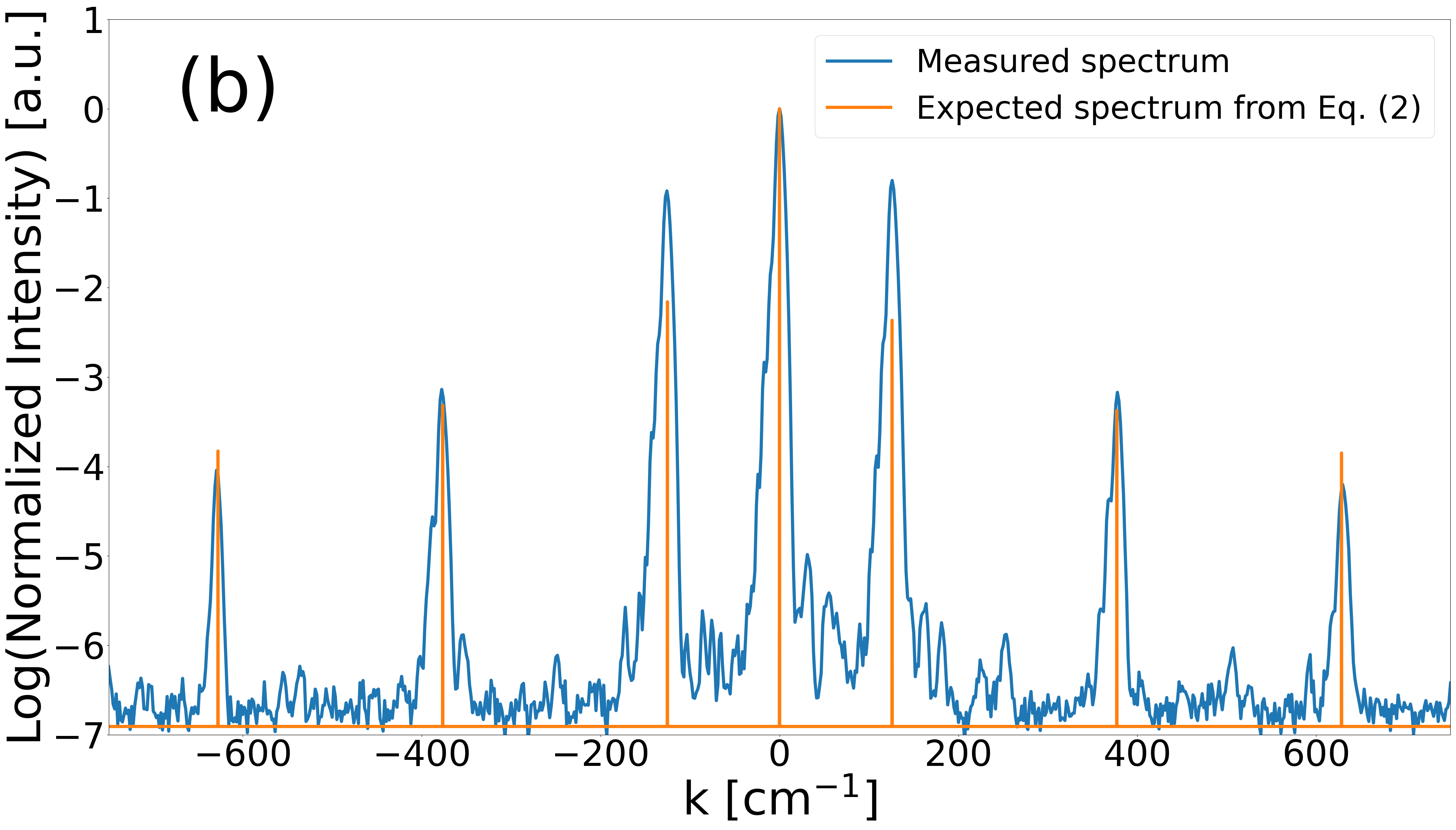} \hspace{0.02\paperwidth}
     \end{tabular}
     \begin{tabular}{cc}
      \includegraphics[width=.4\paperwidth]{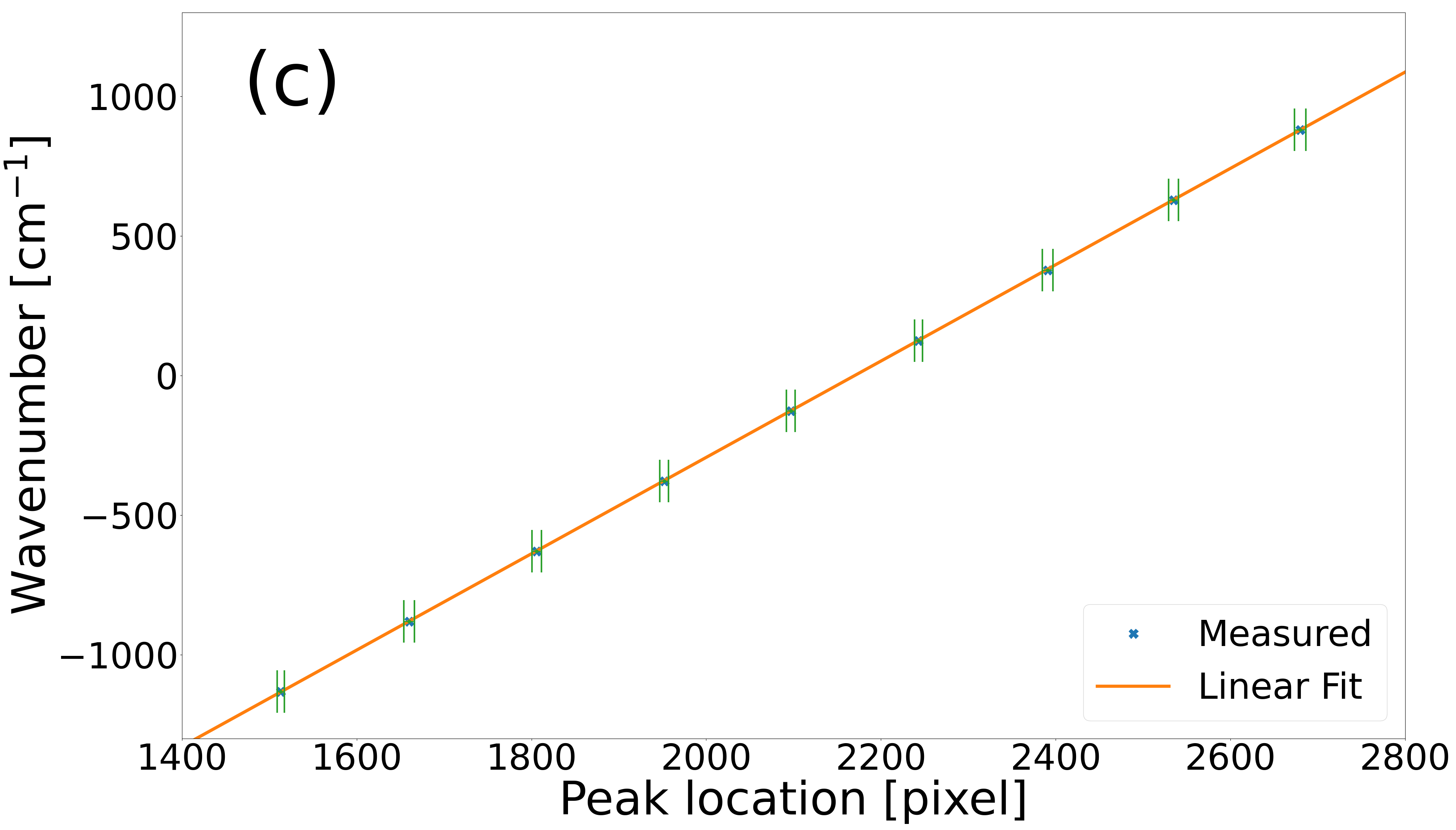} \hspace{0.02\paperwidth}
      \includegraphics[width=.25\paperwidth]{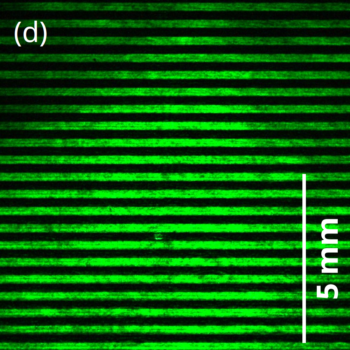} \hspace{0.02\paperwidth}
     \end{tabular}
    \captionsetup{justification=justified,singlelinecheck=false}
     \caption*{\hspace*{\fill}%
        \parbox{1\textwidth}{%
        FIG. 2: (a) - the measured imaging refractometer output with the Ronchi target at the object plane; (b) the lineout from (a) along with the absolute value of \emph{sinc} function from formula (2); (c) peak locations in pixels from the measured spectrum shown in (b) versus the calculated wavenumber value from $k_n=\pi n/a$; (d) the measured shadowgraphy image of the Ronchi target at the object plane.%
        }%
      }
      \label{spec_meas}
    \end{figure*}

\par An attempt can be made to generalize the process described above to a case of turbulent plasma present in the object plane vicinity while removing the infinitely thin density distribution assumption. In that case, the resulting laser field will represent a convolution over the turbulent plasma region. Generally, the optical paths for each of the individual plane waves will become a random variable drawn from the turbulent probability density function. The turbulence level and the average spatial scales will be of general concern, but ultimately the appearance of laser speckle patterns would signal full randomization of the laser field and the need to apply statistical analysis.\cite{Goodman1975} Otherwise, the approach described here may be used.

\par The main goal of this paper is to show that by using a well-defined spatially repetitive target one can calibrate an optical Fourier transformer, as above-described, and allow, in principle, the measurement of the power density spectrum of the laser field. In other words, we can both relate a deflection angle to a wavenumber, as we have shown those are linked, and measure the Fourier space available to the lens used in our setup. To illustrate the described above approach, we define the density distribution at the object plane as shown in Eq. (\ref{def1}). Let \emph{f(x)} be a function that describes the density distribution at the object plane, which in our case is an infinite set of transparent-opaque line pairs with a transparent line width equal to \emph{a} and periodicity \emph{2a}. Eq. (\ref{def1}) forms these line pairs by convolving an infinite set of $\delta$-functions with a unit step function of size $a$. To justify the use of an infinite set, we will assume that during the experiment, the laser beam diameter is smaller than the size of the target, such that its intensity decays to almost zero at the edges. Moreover, we assume that this target satisfies the infinitely thin requirement and, therefore, we can think of \emph{f(x)} as the simplest case of plasma density distribution at the object plane. A schematic view of the optical setup is presented in Fig. \ref{opt_scheme}.
\begin{eqnarray}
    f(x)=\int_{-\infty}^{\infty} \sum_{n=-\infty}^{\infty} \delta(t - 2an)[H((x - t)/a + 1/2)-\nonumber \\
    H((x-t)/a -1/2)] \, dt \label{def1}
\end{eqnarray}
\par Here \emph{H(x)} is a Heaviside step function that represents a square wave of width \textit{a} from $x=-a/2$ to $x=a/2$ and $\delta(t,n)$ is a train of Dirac delta functions (Dirac comb) that has a period of \textit{2a}. We utilize the linear properties of the FT operation, the fact that a Dirac comb function is the FT of itself (up to a scale factor), and that the FT of a square wave function is a \emph{sinc} function, to obtain the following result for the FT of $f(x)$:
\begin{equation}
    g(k) = \frac{a}{\sqrt{2\pi}} \sum_{n=-\infty}^{\infty} \delta\left(k-\frac{\pi n}{a}\right) \times \frac{\sin\left(ka/2\right)}{ka/2} \label{FT}
\end{equation}
\par The function \emph{g(k)} will be zero for all values except for a set of non-zero values that can be found by setting $k_n=\pi n/a$ with $n=0, \pm1, \pm3, \pm5, \pm7 ...$ . Additionally, our target is uniform along the direction of the transparent/opaque lines, allowing the above treatment to be applied to any lineout taken across these line pairs. Figure 2b contains a plot of the magnitude squared of $g(k)$, which corresponds to the intensity of the imaging refractometer image, and clearly indicates that the peak locations follow the previously described dependence with odd Fourier components present along with a $k=0$ component. If the finite size of the target or laser beam is considered, these peaks develop width and there is a structure between the peaks, but the peaks remain centered at the positions indicated above.
\par Thus, if such a gridded target is placed at the object plane of the imaging refractometer device, one can expect to observe a Dirac comb with a spacing of \emph{$2\pi/a$} plus the $k=0$ component in the Fourier space direction (let us refer to it as "y") and continuous narrow lines in the spatial direction (x). This result shows that the laser field stores information about the density variations within a target located at the object plane that can be extracted with an imaging refractometer.
\section{\label{sec:setup}Experimental setup and measurements\protect
}

\par We used an EKSPLA 150 ps, frequency doubled (532 nm) laser with a single vacuum spatial filter ($20 \mu m$ pinhole, $10^{-5}$ torr) for our experiments. The lens arrangement is the same as the one described in Ref. [\onlinecite{hare2021imaging}] and shown in Fig. \ref{opt_scheme}a, with changes only  made to specific focal length values to obtain optimal results, as our experimental vacuum chamber is different. Specifically, we chose $f_1=250mm$ for the spherical lenses and $f_2=500mm$ for the cylindrical lens. For detection, we used the QHY268M-Pro camera, which features a 4210 × 6280 back-illuminated CMOS sensor with a 16-bit depth and very low readout noise. To minimize interference artefacts, we removed the heated glass window used for condensation removal, which consequently disabled the cooling feature. The typical on-sensor integration time was 5 s, while the timescale for the experiment was determined by the laser (150 ps).
\par As described in section \ref{sec:background}, a target having a functional form of \emph{f(x)} placed at the object plane of an imaging refractometer would create a pattern that is spatially uniform in the x-direction and generates narrow peaks with a constant separation of $2\pi/a$ in the Fourier y-direction. These peaks can be labeled in both the negative and positive directions along the \emph{x}-axis, starting from the 1st order (the 0th order should be skipped), and related to corresponding wavenumbers using the relationship from section 2: $k_n=\pi n/a$. Consequently, plotting the \emph{k} values for each peak against its corresponding location in pixels results in a mapping from pixels to \emph{k}'s, as illustrated in Fig. 2. In Fig. 2a, we present the actual output of the imaging refractometer with a Ronchi target having 2 lp/mm ($a=250 \mu m$) at the object plane. Additionally, we include both spatial (horizontal x-direction) and wavenumber (vertical y-direction) scales for the reader's reference. In Fig. 2b, we present a plot of the measured intensity normalized to the $k=0$ value on a log scale along with the normalized $g(k)$ with $a$ set to the value for our target, $a=250 \mu m$. We observe that the peaks align with the locations where \emph{g(k)} has a narrow peak at every $k_na=\pi n$, providing further support for the accuracy of the analytical form. Also, the spacing between the peaks exhibits a highly consistent behavior, as confirmed by the good linear fit results shown in Fig. 2c, for which a normalized $\chi^{2}$ test is close to 1. Nevertheless, slight variations in the spacing on a micrometer scale (2-3 pixels) are present leading to a small shift of the measured peaks. This shift highlights the need for precise optical elements and underscores the sensitivity of the imaging refractometer. In Fig. 2d, we show the measured shadowgraphy image of the same target at the object plane taken with similar laser, which has a non-TEM\textsubscript{00} profile. This image highlights the 1D symmetry in the x-axis of a target and the constant spatial wavelength in the y-axis direction.
\begin{figure*}
    \includegraphics[width=.85\paperwidth]{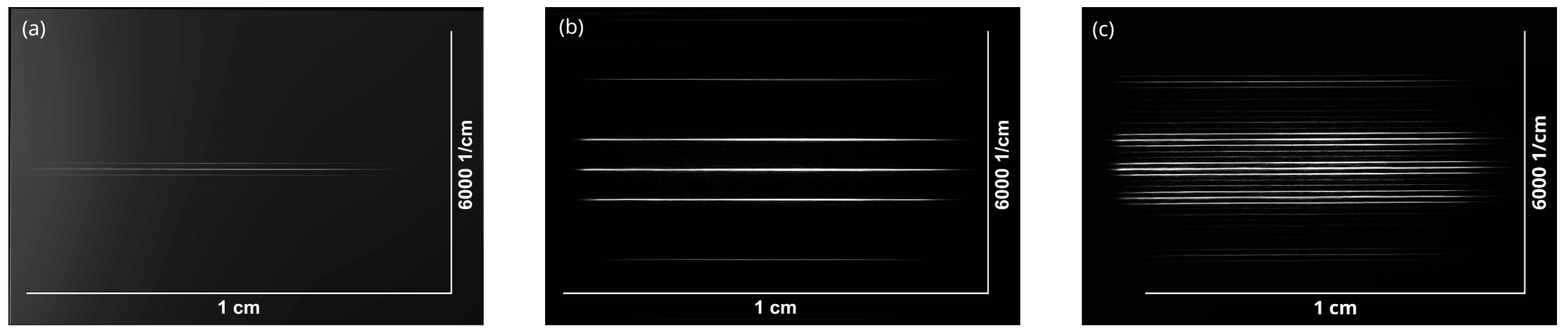} \hspace{0.025\paperwidth}
    \captionsetup{justification=justified,singlelinecheck=false}
 \caption*{\hspace*{\fill}%
    \parbox{1\textwidth}{%
    FIG. 3: (a) - the measured imaging refractometer output with the Ronchi 2 lp/mm target at the object plane; (b) - the same, but 10 lp/mm Ronchi target; (c) 10 lp/mm Ronchi target stacked back-to-back with 2 lp/mm Ronchi.%
    }%
}
\end{figure*}
\par We note here that the refractive index distribution at the object plane (of the target) in this particular case has only two distinct values, namely $n=0$ and $n=1$. However, due to the fact that an imaging refractometer is insensitive to absolute values, but to the variation in the refractive index, the resultant measurement in the Fourier direction (y-axis) will be the same even for cases where this variation, $\delta n$, is small.
\par In Fig. 3 we show how an imaging refractometer reacts to different density patterns along the probing beam path where at the object plane we place 2 lp/mm in Fig. 3(a), 10 lp/mm in Fig. 3(b), and both targets with the etched sides facing each other in Fig. 3(c). As expected, the laser intensity distribution after the first target (10 lp/mm Ronchi) behaves as a convolution while propagating through the next one, which is also a manifestation of the linear response of the system. Therefore, any departure from the initial assumption of infinitely thin plasmas would result in the wavenumber line bifurcation. The intensity of the bifurcated lines will drop as fast as \emph{sin(x)/x} and its width can be estimated from the first order spacing, given it is resolvable, that is $\pi/\delta a_1$, where $\delta a_1$ is the average spacing along the laser beam propagation path. In our case, it is the width of a transparent line of the second target, 250 $\mu m$. Furthermore, a conclusion can be made that the measured wavenumber intensity, although convoluted with different structures and bifurcated, will reflect the actual plasma density distribution. Specifically, if certain wavenumbers are part of the plasma density distribution, these wavenumbers will be measured with an imaging refractometer, which is seen in Fig. 3c. Additionally, the lines position and spread remained the same in Figs. 3(a-c), which means that the Fourier space size and calibration is not altered by the convolution, which will ease post-processing, including the analysis of the secondary structures. These structures (additional lines) are due to the perturbation of the incoming single-mode laser profile by the first target. To identify the secondary structures, which may constitute for example Rayleigh-Taylor bubbles and/or spikes, it is useful to have additional diagnostics, e.g. shadowgraphy, and numerical capabilities, for example, Beam Propagation Method\cite{okamoto2006} type simulations and different magnetohydrodynamical codes.\cite{Seyler_Perseus}

\section{\label{sec:summary}Summary\protect
}
\par In this paper, we demonstrate the procedure to calibrate an imaging refractometer by placing the one-dimensionally uniform, spatially repetitive target at its object plane and measuring the pixel-to-k ratio. This calibration procedure demonstrates how to interpret the measured laser field intensity per unit wavenumber per unit solid angle. While we used a Ronchi target in our experiment to provide this calibration, any target with an analytic FT is suitable for this purpose. As a result of this calibration, the optical arrangement described here is capable of detecting wavenumbers from $k\approx 1cm^{-1}$ to $k\approx 3000cm^{-1}$, which amounts to density fluctuation wavelengths less than 1 cm down to approximately 20 µm, demonstrating its remarkable dynamic range. By shifting the $k=0$ line to the bottom of the detector screen, the maximal \emph{k} can be as high as $6000cm^{-1}$. In addition, a second imaging refractometer leg can be added to the optical setup with its cylindrical lens rotated 90 degrees relative to the first one. This arrangement leads to the ability to measure wavenumber spectra in two dimensions simultaneously while having two-dimensional spatial information as well. 
\par We conclude that the measurements acquired with an imaging refractometer, which doesn't feature laser speckle patterns, will show a true spatial spectrum width and will reflect the plasma dispersion at the object plane. Otherwise, the statistical analysis is required. Possible examples for the use of an imaging refractometer device may include an imploding shock in cylindrical geometry with a relatively thin turbulent layer behind it, a shockwave bouncing off of the wall, and a turbulent layer produced by colliding jets. 

\begin{acknowledgments}
The authors would like to thank Dr. S. Merlini, Dr. S. N. Bland, and Prof. S. V. Lebedev for fruitful discussions, and Todd Blanchard, Daniel Hawkes, and Harry Wilhelm for their support in conducting this experiment. This research was supported by the Cornell Laboratory of Plasma Studies, by the Engineering Dean's office through the College Research Incentive Program, by the K. Bingham Cady Memorial Fund, and by the Air Force Office of Scientific Research under award number FA9550-24-1-0066.
\end{acknowledgments}

\section*{Data Availability Statement}
The data that support the findings of this study are available from the corresponding author upon reasonable request.
\nocite{*}
\bibliography{imref_refs}

\end{document}